\documentclass{emulateapj}

\usepackage{epsfig}

\begin{document}

\title{The $^{7}$B\lowercase{e}(\lowercase{d},\lowercase{p})2$\alpha$ cross section at Big Bang energies and the primordial $^7$L\lowercase{i} abundance}

\author{C. Angulo, 
E. Casarejos, 
M. Couder\footnote{Present address: Department of Physics, University of Notre Dame, IN 46556-5670, USA}, 
P. Demaret,
P. Leleux\footnote{Directeur de Recherches F.N.R.S., Belgium}, 
F. Vanderbist}
\affil{Centre de Recherches du Cyclotron and Institut de Physique Nucl\'eaire, 
UCL, \\
B-1348 Louvain-la-Neuve, Belgium; angulo@cyc.ucl.ac.be, enrique@cyc.ucl.ac.be, couder@fynu.ucl.c.be, leleux@fynu.ucl.ac.be, vanderbist@fynu.ucl.ac.be}
\author{A. Coc, J. Kiener, V. Tatischeff}
\affil{CSNSM, CNRS/IN2P3/UPS, B\^at. 104, F-91405 Orsay Campus, France; coc@csnsm.in2p3.fr, kiener@csnsm.in2p3.fr, tatischeff@csnms.in2p3.fr}
\author{T. Davinson, A.S. Murphy}
\affil{School of Physics, The University of Edinburgh, Edinburgh EH9 3JZ, UK; td@ph.ed.ac.uk, amurphy@ph.ed.ac.uk}
\author{N.L. Achouri, N.A. Orr}
\affil{LPC, ENSICAEN and Universit\'e de Caen, IN2P3-CNRS, Caen Cedex, France; achouri@lpccaen.in2p3.fr, orr@lpccaen.in2p3.fr}
\author{D. Cortina-Gil}
\affil{Depto.~de F{\'i}sica de Part{\'i}culas, Universidad de Santiago de Compostela, Spain; d.cortina@usc.es}
\author{P. Figuera}
\affil{INFN-Laboratori Nazionali del Sud, Catania, Italy; figuera@lns.infn.it}
\author{B.R. Fulton}
\affil{Department of Physics, University of York, York YO10 5DD, UK; brf2@york.ac.uk} 
\author{I. Mukha\footnote{On leave from RRC Kurchatov Institute, RU-123481 Moscow, Russia.}}
\affil{Kern- en Stralingsfysica, Katholieke Universiteit Leuven, Leuven, Belgium; imukha@gsi.de}
\author{E. Vangioni}
\affil{Institut d'Astrophysique de Paris, 98$^{\mathrm bis}$ Boulevard Arago,
F-75014 Paris, France; vangioni@iap.fr}

\begin{abstract}
The WMAP satellite, devoted to the observations of the anisotropies of the 
Cosmic Microwave Background (CMB) radiation, has recently provided a 
determination of the baryonic density of the Universe with unprecedented 
precision. 
Using this, Big Bang Nucleosynthesis (BBN) calculations predict a primordial 
$^7$Li abundance which is a factor $2-3$ higher than that observed in galactic halo dwarf stars. 
It has been argued that this discrepancy could be resolved if the 
$^{7}$Be(d,p)2$\alpha$ reaction rate is around a factor of 100 larger than has 
previously been considered.
We have now studied this reaction, for the first time at energies
appropriate to the Big Bang environment, at the CYCLONE 
radioactive beam facility at Louvain-la-Neuve. The cross section was found to be 
a factor of 10 {\em smaller} than derived from earlier measurements.
It is concluded therefore that nuclear uncertainties cannot explain the discrepancy between
observed and predicted primordial $^7$Li abundances, and an alternative 
astrophysical solution must be investigated.

\end{abstract}


\section{Introduction}

Using the WMAP-determination of the baryonic density \citep{bennett03,spergel03}, one
obtains predictions of the abundances of the light element isotopes produced in 
Big Bang Nucleosynthesis \citep{cyburt03,coc02,coc04}. 
While the overall values from theoretical predictions and from the 
observational determinations of the abundances of D and $^4$He are in good agreement, 
the theory tends to predict a higher $^7$Li abundance (by a factor 2 to 3) than is observed 
in the atmospheres of halo dwarf stars \citep{ryan00}.
The NACRE compilation \citep{nacre} provided a new set of reaction 
rates that were used to update the predictions of contemporary 
Big Bang Nucleosynthesis (BBN) \citep{vangioni00}. 
At that time, the baryonic densities obtained from CMB observations on the 
one hand and comparison between BBN calculations and spectroscopic data
on the other hand were only marginally compatible \citep{coc02}. 
In order to improve the nuclear network, \citet{descouvemont04} 
recently performed a re-analysis of low energy data from the 10 key nuclear reactions involved
in BBN, by using R-matrix theory \citep{lane58} and evaluating the remaining uncertainties in a statistically robust
formalism.
Using this improved network, \citet{coc04} have recently calculated BBN light element 
productions assuming for the baryonic density the very precise value provided
by WMAP \citep{spergel03} and obtained 
$^{7}$Li/H$ = 4.15^{+0.49}_{-0.45}\times10^{-10}$ compared to the observed value
Li/H $\simeq 1-2\times10^{-10}$, confirming the $^{7}$Li 
discrepancy. 

However, it has been shown \citep{coc04} that the $^{7}$Be(d,p)2$\alpha$ reaction (which 
destroys the $^{7}$Be that is the source of $^{7}$Li at high baryonic density), would
solve the $^{7}$Li problem {\em if} its cross section were much higher
than assumed.
Importantly, prior to the present work, {\em no direct experimental data at BBN energies} 
were available (for $T = 0.5 - 1$ GK, the Gamow window is $E = 0.11 - 0.56$ 
MeV).
In fact, the $^{7}$Be(d,p)2$\alpha$ reaction rate relied on an extrapolation made by 
\citet{parker72} based on experimental data at center-of-mass (c.m.) energies of 0.6 to 1.3 MeV from 
\citet{kavanagh60}. In this experiment, protons corresponding to the $^8$Be 
$0^+$ ground state (g.s.) and first excited state (3.03 MeV, $2^+$) 
were detected at 90$^{\circ}$ using a NaI(Tl) detector. 
Assuming an isotropic angular distribution, \citet{parker72} multiplied the measured differential cross 
section by $4 \pi$ and by a further factor of 3 to take into account the estimated
contribution of the higher energy $^8$Be states, not observed by 
\citet{kavanagh60}. Consequently, a
constant $S$-factor of 100 MeV-barn was adopted.

In order to obtain $^7$Be(d,p)2$\alpha$ reaction cross section at BBN energies, 
we have performed an experiment at the CYCLONE radioactive beam facility at Louvain-la-Neuve, Belgium, using
an isobarically pure $^7$Be radioactive beam. 
The experimental method and results are presented in Section 2. The astrophysical consequences are discussed in Section 3. 
The conclusions are given in Section 4.

\section{Experimental method and results}

\begin{figure}[ht]
\begin{center}
\end{center}
\caption{Schematic view of the experimental set-up.}
\label{fig1}
\end{figure}

The measurements were performed using a post-accelerated $^{7}$Be$^{1+}$ radioactive beam at a nominal energy of 5.8 MeV 
provided by the CYCLONE110 cyclotron.
A detailed description of the production of the $^{7}$Be beam can be found 
in \citet{gaelens03}.
To suppress the contamination from the $^7$Li isobaric beam, the $^7$Be beam 
was completely stripped to $^7$Be$^{4+}$ by transmission through a thin $^{12}$C foil, 
prior to analysis by a dipole magnet. 
Prior to the $^{7}$Be(d,p)2$\alpha$ measurement, the beam energy was determined using
a calibrated Si detector situated at 0$^{\circ}$. A laboratory energy of 
5.55 MeV (FWHM $\sim 4\%$) was determined, including a correction for pulse height defect. 
This energy was degraded to 1.71 MeV (FWHM $\sim 12\%$) 
using a 6 $\mu$m Mylar foil located at 50 cm upstream of the target. 
No $^7$Li 
contamination was observed, consistent
with lithium isotopes being unable to support a 4+ charge state. 
The target consisted of a 200 $\mu$g/cm$^2$ (CD$_2$)$_n$ self-supporting foil. 
With this set-up, we were able to investigate the center-of-mass energy range between 
1.00 and 1.23 MeV (for a beam energy of 5.55 MeV, without degrader) and between 0.13 and 0.38 MeV (for 1.71 MeV, with degrader). 
The cross section measurement was averaged over these energy ranges.
In addition to the feeding of the ground and first excited states of $^8$Be \citep{kavanagh60}, 
we were able to observe the $^7$Be+d reaction via other kinematically allowed higher energy levels, 
mainly through a very broad $4^+$ state ($\Gamma \simeq 3.5$ MeV) situated at an excitation energy of 11.4 MeV in $^8$Be \citep{tilley04}. 
At the beam energy of 5.55 MeV, 
several states in $^8$Be above the $^7$Be+d-p threshold are present but due to the Coulomb barrier in the final state, 
their contribution are expected to be negligible.
The $Q$ value of the $^7$Be(d,p)$^8$Be reaction is 16.49 MeV, thus the laboratory energies of 
protons and $\alpha$ particles are high.
For example, a 5.55 MeV $^7$Be beam traversing a 200 $\mu$g/cm$^2$ (CD$_2$)$_n$ target will lead to the
production of protons with energies anywhere between about 7.5 and 22 MeV, for the range of angles covered.
Thus, to clearly identify the protons from the $^7$Be+d reaction from those arising from reactions on the
C content of the target, a stack of two `LEDA' silicon strip detector arrays \citep{davinson00} were employed covering a laboratory angular range of $\theta_{\rm lab} = 7.6^{\circ} - 17.4^{\circ}$. 
A schematic view of the
experimental set-up is shown in Figure 1. The $\Delta E_1$ detector consisted of eight sectors of 0.3 mm 
thickness, while the $\Delta E_2$ detector consisted of four sectors of 0.3 mm thickness and four 
of 0.5 mm thickness. They were calibrated using a 3-line $\alpha$-source ($^{239}$Pu, $^{241}$Am, $^{244}$Cm) and a precision pulser.
This $\Delta E_1 - \Delta E_2$ detector system allowed a clear identification of the protons produced in the $^7$Be(d,p)$^8$Be reaction.
All the particles that are not stopped
in the front $\Delta E_1$ detector and that are either stopped or that left energy on the back $\Delta E_2$ detector
are protons (having an energy of more than 6.5 MeV) that populated levels up to the 11.4 MeV state in $^8$Be. 
We were able to measure the $^{7}$Be(d,p)2$\alpha$ cross section 
up to an excitation energy in $^8$Be of $E_x = 13.8$ MeV for a beam energy 
of 5.55 MeV and of $E_x = 11.5$ MeV for 1.71 MeV. Only about 50\% of the contribution of the $4^+$ broad state was observed at 1.71 MeV.
Other light particles (p, d, $^3$He, $^4$He)
from $^7$Be+$^{12}$C reactions, as well as recoils and scattered particles, were completely stopped in $\Delta E_1$. 

Figures 2 and 3 show spectra obtained at beam 
energies of 5.55 and 1.71 MeV, respectively.
The spectrum obtained at 1.71 MeV (Figure 3) was accumulated over about 26 hours of 
running time with 
an averaged $^7$Be beam intensity of $2\times10^6$ pps. 
As can be seen, the proton signals are well separated from the background 
signals ($\Delta E_2 < 1$ MeV), which are produced by random coincidences of 
$\alpha$ particles, scattered $^7$Be and recoil ions in the $\Delta E_1$ 
detector with $\beta$ particles in $\Delta E_2$.
The locus with negative slope contains protons that have passed through
the front $\Delta E_1$ detector and stopped in the back $\Delta E_2$ detector.  The two
loci with positive slope are events in which the proton has sufficient
energy to pass through both detectors completely. There are two bands
because of the two different thickness $\Delta E_2$ detectors. The most strongly populated regions
at the lower left of these bands correspond to protons losing the least
energy in passing through the silicon, and thus to the highest energy
events. By considering the kinematics and energy losses in silicon \citep{srim03}, 
together with the straggling of the beam and experimental energy
resolution, one may then identify events on the positive slope locus up to 2.5 (3.9) MeV for the 0.3
(0.5) mm Si thickness wafer, as corresponding to events in which the
recoiling $^8$Be nucleus is in either the $0^+$ g.s. or the $2^+$ excited state
(the energy resolution is insufficient to resolve the two).
The total statistical error was 10\% for the beam energy of 1.71 MeV and less than 2\% for 5.55 MeV
(for protons populating the $0^+$ and $2^+$ states it was 13\% and 2.5\%, respectively).
The absolute normalization was obtained using events arising from the elastic 
scattering of the $^7$Be on the C content of the target (as recorded by the $\Delta E_1$ detector in which the $^7$Be are stopped), and 
assuming that the $^7$Be+$^{12}$C elastic scattering follows the Rutherford law. This assumption is realistic at energies below the Coulomb barrier, as is the case here. 

To calculate the average cross section over the energy ranges and angular coverages $(d \overline{\sigma} / d \Omega)$, the number of counts was corrected for 
the detector solid angle (uncertainty $\pm 5\%$), the number of deuterons in the target ($\pm 10\%$) and
the total number of incoming beam particles ($\pm7\%$ and $\pm26\%$ at the higher and lower beam energies, respectively), and transformed into the c.m. system.

\begin{figure}[h]
\begin{center}
\epsfig{file=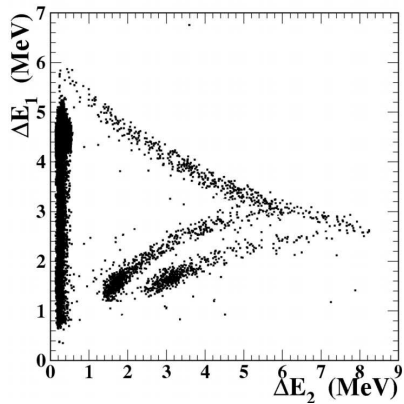, height=5.4cm}
\end{center}
\caption{$\Delta E_1$-$\Delta E_2$ spectrum at 
a beam energy of 5.55 MeV on a 200 $\mu$gr/cm$^2$ (CD$_2$)$_n$ target. The c.m.~energy range covered
is 1.0 to 1.23 MeV.}
\label{fig2}
\end{figure}

\begin{figure}[h]
\begin{center}
\epsfig{file=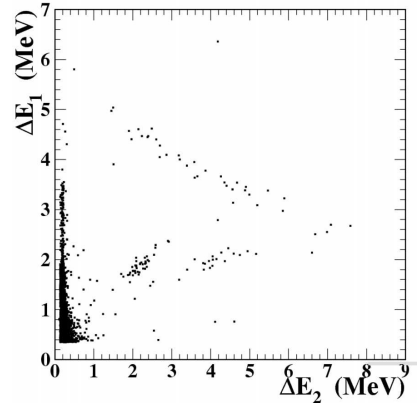, height=5.4cm}
\end{center}
\caption{Same as Fig.~3 for a beam energy of 1.71 MeV, corresponding to a c.m.~energy range 
of 0.13 to 0.38 MeV.}
\label{fig3}
\end{figure}

The proton angular distribution over the angular range covered here was found to be isotropic at both energies.
Thus, we assumed full
isotropy and calculated the average cross sections, 
$\overline{\sigma} = 7.5 \pm 0.8(stat) \pm 2.6(sys)$ mb at the effective energy of 0.37 MeV and  
$\overline{\sigma} = 386 \pm 7(stat) \pm 50(sys)$ mb at 1.15 MeV. 
The summed contribution of the $0^+$ and $2^+$ states was about
64\% of $\overline{\sigma}$ at 1.15 MeV. At 0.37 MeV, $\overline{\sigma}$ includes the contribution 
of the ground and $2^+$ states and about 50\% of that of the
$4^+$ broad state. Due to the low penetration probability
($\ell=4$), the contribution of the $4^+$ state should be less than 36\%. Thus, the $\overline{\sigma}$ value 
at 0.37 MeV corresponded to more than 80\% of the total cross section 
(for a $4^+$ state with $\Gamma \simeq 3.5$ MeV). This was taken 
into account in the systematic uncertainty.

In nuclear astrophysics it is usual to present the cross section in the form of the astrophysical $S$-factor $S(E)$ given by \citep{clayton83},
\begin{equation}
S(E) = \sigma(E) \exp(2 \pi \eta) E,
\end{equation}
where $\eta$ is the Sommerfeld parameter ($\eta = Z_1Z_2 e^2/\hbar v$, 
with $Z_1$ and $Z_2$ the charge numbers of the target and beam and $v$ is the velocity) 
and $E$ is the effective c.m.~energy. In the absence of sharp resonances, the $S$-factor varies smoothly with energy.
Figure 4 shows the $^7$Be(d,p)$^8$Be astrophysical $S$-factor $\overline{S}(E)$ in MeV-barn as a function of the c.m.~energy. 
For a comparison with the data of \citet{kavanagh60} (open circles), the present data 
including only contributions from the ground and first excited states of $^8$Be (filled circles) are shown. The agreement with the \citet{kavanagh60} data 
at overlapping energies is satisfactory, given the systematic uncertainties. 
The total S-factor is also shown (filled triangles). 
The present data show that the higher energy states not observed by \citet{kavanagh60} contribute about 35\% of the total $S$-factor instead of the 
300\% estimated by \citet{parker72}. Hence, the $^7$Be(d,p)$^8$Be reaction rate is 
smaller by a factor of about 2 at energies in the range 1.0 to 1.23 MeV and 
by about 10 at energies relevant to BBN, than previously estimated.
This excludes a nuclear solution to the primordial lithium abundance problem via the 
$^7$Be(d,p)$^8$Be reaction as its effect is completely negligible compared to the 
7\% ($1 \sigma$) nuclear uncertainty on the $^7$Li yield. Nevertheless, these results allow a 
more accurate determination of the $^7$Li abundance using BBN models.

\begin{figure}[h]
\begin{center}
\epsfig{file=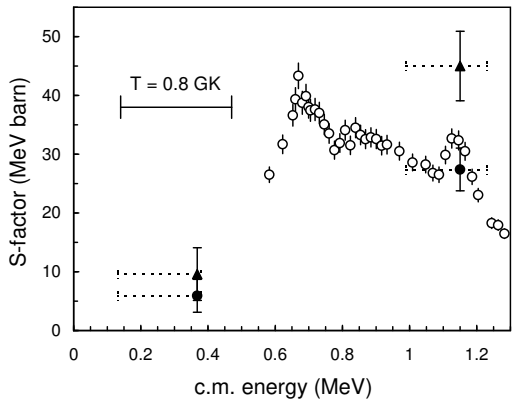, height=5.7cm}
\end{center}
\caption{Astrophysical $S$ factor of the $^7$Be(d,p)$^8$Be reaction. Open circles: data from \citet{kavanagh60}; filled circles: 
present data including contributions from the ground and first excited states of $^8$Be only; filled triangles: total $S$-factor derived from the present experiment. 
The vertical error bars are the total error. The horizontal dotted bars indicate the energy range covered at each data point.
The Gamow window for a typical BBN temperature T=0.8 GK is also shown.}
\label{fig3}
\end{figure}

\section{Astrophysical consequences}

Since the pioneering work of \citet{spite82}, who found a value of
Li/H$\approx 1.2 \times 10^{-10}$ independent of Fe/H (for [Fe/H]$<-1.3$) 
there have been many independent observations of Li 
confirming the existence of a plateau and suggesting that this abundance 
reflects the primordial Li value. 
However, the Li abundance extracted from observations depends drastically 
on the assumed 
surface temperature of the star (\citet{fields05}).
Recent observations \citep{ryan00} have lead to 
Li/H $= (1.23^{+0.68}_{-0.32})\times10^{-10}$ which is very close to the first evaluation \citep{spite82}. 
The more recent work studied and quantified the various sources of uncertainty: 
extrapolation, stellar depletion and stellar atmosphere parameters. 
Compared to the WMAP+BBN value, the discrepancy is a factor of $\sim 3.4$. 
If it is shown that there is a mechanism by which the outer layers of
Population II stars are transported deep into the stellar interior, then
there are several ways in which Li abundances might be depleted over the
lifetime of the star. In this context,
the current estimates for possible depletion factors may be in the range
$\sim0.2-0.4$ dex \citep{vauclair98,pinsonneault02,richard04}. 
However, the data typically show negligible intrinsic spread in the Li abundance leading to the conclusion
that depletion in these stars is of the order of 0.1 dex. 

Recently, \citet{melendez04} have obtained a higher value for the Li plateau abundance
($2.34\times 10^{-10}$) 
due to a new effective temperature scale which is higher at low metallicity. 
This new evaluation diminishes the discrepancy, without canceling it.
The observation of $^6$Li is also of interest, since, because it is more 
fragile than $^7$Li, it can provide yet more severe constraints upon possible
depletion mechanisms \citep{lambert04,rollinde05}. Finally, in spite of the various 
uncertainties related to Li observations and to the stellar models, it is 
very difficult to reconcile the BBN $^7$Li and the Spite plateau which 
presents a narrow dispersion all along the metallicity scale.

\section{Conclusions}

The existence of the Spite plateau for Li seems to indicate that low 
metallicity halo stars are indeed representative of the primordial 
BBN abundance. In particular, the isotope
$^7$Li plays a key role as a bridge between Big Bang Nucleosynthesis, stellar 
evolution and galactic cosmic-ray nucleosynthesis.  
At present there is a significant discrepancy between the BBN-predicted 
$^7$Li abundance (assuming a baryon density consistent with the concordance 
model derived from observations of anisotropies in the microwave background) 
and the abundance determined from the observations of Li in the 
atmospheres of halo stars.
The experiment reported here demonstrates that the $^{7}$Be(d,p)2$\alpha$  
$S$-factor at BBN energies 
was not underestimated by \citet{parker72} but, on the contrary, 
{\em overestimated}.
The discrepancy cannot therefore be resolved by nuclear physics inputs to BBN calculations. 
The remaining conventional options (those not invoking physics beyond the 
Standard Model) are an adjustment of the stellar input parameters needed to 
extract the Li abundances from observations, or stellar depletion of $^7$Li.
However, models must be
constructed to avoid dispersion in the $^7$Li abundances over a wide range of 
stellar parameters, which is a real challenge.
The origin of the discrepancy in the Li abundance remains a 
challenging issue.

\section*{Acknowledgments}
This work was partially supported by the EC under contract no.~HPRI-CT-1999-00110, the Belgian Inter-University Attraction Poles P5/07, the IN2P3 (France), PICS 1076 USA/CNRS, and
the UK EPSRC. EV is grateful to K. Olive for fruitful discussions.

\end{document}